\documentstyle[aps,prl,floats,epsf]{revtex}
\tighten
\draft
\begin{document}

\def\nue {{\nu_e}}
\def\numu {{\nu_\mu}}
\def\nutau {{\nu_\tau}}
\def\nube {{\bar{\nu}_e}}
\def\nubmu {{\bar{\nu}_\mu}}
\def\nubtau {{\bar{\nu}_\tau}}

\twocolumn[\hsize\textwidth\columnwidth\hsize\csname
@twocolumnfalse\endcsname

\title{Enhanced signal of astrophysical tau neutrinos 
propagating through Earth}
\author{\mbox{John F. Beacom,$^{1}$} 
\mbox{Patrick Crotty,$^{2}$}
and \mbox{Edward W. Kolb$^{1,3,4}$}}
\address{\mbox{$^1$ NASA/Fermilab Astrophysics Center, 
Fermi National Accelerator Laboratory, Batavia, Illinois 60510-0500, USA}
\mbox{$^2$ Department of Physics, University of Chicago,
Chicago, Illinois 60637, USA}
\mbox{$^3$ Department of Astronomy and Astrophysics,
University of Chicago, Chicago, Illinois 60637, USA}
\mbox{$^4$ TH Division, CERN, CH-1211 Geneva 23, Switzerland}
\\
{\tt beacom@fnal.gov},
{\tt prcrotty@oddjob.uchicago.edu},
{\tt rocky@fnal.gov}}
\date{November 26, 2001}
\maketitle

\begin{abstract}
Earth absorbs $\nue$ and $\numu$ of energies above about 100 TeV.  As
is well-known, although $\nutau$ will also disappear through charged-current
interactions, the $\nutau$ flux will be regenerated by prompt tau
decays.  We show that this process also produces relatively large
fluxes of secondary $\nube$ and $\nubmu$, greatly enhancing the
detectability of the initial $\nutau$.  This is particularly important
because at these energies $\nutau$ is a significant fraction of the
expected astrophysical neutrino flux, and only a tiny portion of the
atmospheric neutrino flux.
\end{abstract}

]
\narrowtext

%%%%%%%%%%%%%%%%%%%%%%%%%%%%%%%%%%%%%%%%%%%%%%%%%%%%%%%%%%%%%%%%%%%%%%
%%%%%%%%%%%%%%%%%%%%%%%%%%%%%%%%%%%%%%%%%%%%%%%%%%%%%%%%%%%%%%%%%%%%%%
%\section{Introduction}
%%%%%%%%%%%%%%%%%%%%%%%%%%%%%%%%%%%%%%%%%%%%%%%%%%%%%%%%%%%%%%%%%%%%%%
%%%%%%%%%%%%%%%%%%%%%%%%%%%%%%%%%%%%%%%%%%%%%%%%%%%%%%%%%%%%%%%%%%%%%%

{\it Introduction.---} 
A variety of astrophysical and exotic sources are expected to produce
large fluxes of ultra-high-energy particles, including photons,
protons, and neutrinos.  Sources such as active galactic nuclei and
gamma-ray bursts are at high redshifts, and the photons and protons
will be attenuated by scattering on cosmic radiation backgrounds.  The
protons may also be deflected by magnetic fields.  However, neutrinos
from distant sources will neither be attenuated nor deflected, thus
allowing true neutrino astronomy of the high-redshift universe.
Further, it is now known that about one-third of the energy density of
the universe is accounted for by particles outside the
Standard Model.  Among the possibilities are variants of superheavy
dark matter, which may produce ultra-high-energy particles from their
annihilations or decays.  Such signals would be most readily observed
from local sources.  The existence of cosmic rays with energies up to
and above $10^{20}$ eV is a powerful argument for the existence of
ultra-high-energy astrophysical neutrinos.  With the upcoming km-scale
detectors, a new era of neutrino astronomy will begin, and the
energies, directions, and flavors of astrophysical neutrinos will
provide important clues to the most violent astrophysical objects and
the nature of particle dark matter\cite{Gaisser,Learned}.

While these detectors can suppress downgoing atmospheric muons,
atmospheric neutrinos are a much more challenging background.  Like
astrophysical neutrinos, these can pass through Earth and create
upgoing muons (for example).  They are produced in the decays of
secondaries produced by the collisions of cosmic-ray protons with
Earth's atmosphere.  The decays $\pi^+/K^+ \rightarrow \mu^+ \numu$
and $\mu^+ \rightarrow e^+ \nue \nubmu$ and their charge conjugates
produce fluxes in the flavor ratio $\nue:\numu:\nutau \simeq 1:2:0$,
where neutrinos and antineutrinos have been combined.  These ratios
vary with energy and angle when the production of other mesons and
their decay lengths relative to the slant depth of the atmosphere are
taken into account.  In particular, at the highest neutrino energies,
there is a tiny relative $\nutau$ flux from charm
decay~\cite{Pasquali}.  The atmospheric neutrino flux falls roughly as
$E^{-3}$, whereas general considerations for astrophysical neutrino
fluxes predict shallower slopes, and hence that they will dominate
above some energy.  For diffuse fluxes, this crossover is thought to
be at least 10 TeV (point and/or transient sources can be identified
at lower energies)~\cite{Gaisser,Learned}.

Neutrino oscillations have an important consequence for suppressing
the atmospheric neutrino background.  The Super-Kamiokande
atmospheric neutrino results~\cite{SK} strongly favor $\numu
\leftrightarrow \nutau$ oscillations with $\sin^2{2\theta} \simeq 1$
and $\Delta m^2 \simeq 3 \times 10^{-3}$ eV$^2$.  At the high energies
considered here, $\numu \leftrightarrow \nutau$ oscillations will {\it
never} occur for atmospheric neutrinos, and their flavor ratios will
reflect the production mechanism only.  In the astrophysical
production scenarios, the flavor ratios are also thought to be $1:2:0$
(again combining neutrinos and antineutrinos), reflecting their
production by ultra-high-energy protons.  But for the
astrophysical sources, $\numu \leftrightarrow \nutau$ mixing will {\it
always} be complete because of the long path lengths, and so these
ratios become $1:1:1$.  In some exotic models~\cite{AHK,MacGibbon},
the initial $\nutau$ flux is large; in any case, a large $\nutau$
component in the astrophysical neutrino flux is guaranteed.

This simple fact is very important for the detection of astrophysical
neutrinos above the atmospheric neutrino background.  This is because
$\nue$ and $\numu$ above about 100 TeV will be absorbed by Earth,
whereas the absorbed $\nutau$ flux will be regenerated by prompt tau
decays~\cite{Halzen}.  We point out a new effect: that the
regeneration of the $\nutau$ flux also creates relatively large
secondary fluxes of $\nube$ and $\nubmu$.  These make the detection of
astrophysical sources significantly {\it easier}, and must be accounted
for to properly deduce the source spectra and flavor composition.

%%%%%%%%%%%%%%%%%%%%%%%%%%%%%%%%%%%%%%%%%%%%%%%%%%%%%%%%%%%%%%%%%%%%%%
%%%%%%%%%%%%%%%%%%%%%%%%%%%%%%%%%%%%%%%%%%%%%%%%%%%%%%%%%%%%%%%%%%%%%%
%\section{Neutrino propagation in Earth}
%%%%%%%%%%%%%%%%%%%%%%%%%%%%%%%%%%%%%%%%%%%%%%%%%%%%%%%%%%%%%%%%%%%%%%
%%%%%%%%%%%%%%%%%%%%%%%%%%%%%%%%%%%%%%%%%%%%%%%%%%%%%%%%%%%%%%%%%%%%%%

\begin{figure}[t]
\centerline{\epsfxsize=3.5in \epsfbox{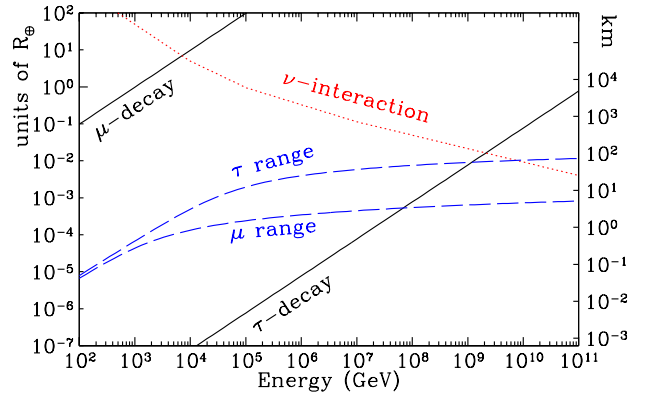}}
\caption{Some important length scales, in units of Earth radii and km,
versus the energy in GeV for the neutrino or charged lepton.  {\it
Solid lines:} the $\mu$ and $\tau$ decay lengths, i.e., the distances
over which a fraction $e^{-1}$ of the the initial fluxes would survive
(ignoring energy loss).  {\it Dashed lines:} the $\mu$ and $\tau$
ranges, i.e., the distances in standard rock ($22$ g mol$^{-1}$) at a
density 8 g cm$^{-3}$ over which the which they would be {\it fully}
stopped by their electromagnetic interactions (ignoring decays).  {\it
Dotted line:} the neutrino interaction length (the same for all
flavors), i.e., the distance at a density 8 g cm$^{-3}$ over which a
fraction $e^{-1}$ of the initial flux would survive.  With the
exception of the Glashow resonance for $\nube$ at 6.3 PeV, this figure
is the same for antiparticles.  While the density varies significantly
with nadir angle, 8 g cm$^{-3}$ is representative.}
\label{fig:lengths}
\end{figure}

{\it Neutrino propagation in Earth.---}
Above about 100 TeV, all neutrino flavors have a high probability of
interacting in Earth via neutrino-nucleon scattering, with a total
cross section $\sigma \propto E^{0.5}$~\cite{GQRS}.  Except for $\nube
+ e^- \rightarrow W^-$ at the Glashow resonance (6.3 PeV),
neutrino-electron scattering is irrelevant.  Charged-current (CC) and
neutral-current (NC) interactions on nucleons occur in the ratio
$0.71:0.29$, and in both cases, the surviving lepton carries about
$75\%$ of the initial neutrino energy~\cite{GQRS}.  The cross sections
for neutrinos and antineutrinos are approximately equal.  The neutrino
interaction length is shown in Fig.~\ref{fig:lengths}; it is the same
for all flavors since even tau mass threshold effects are negligible.

In the more likely CC interaction, the critical question is the
subsequent fate of the charged lepton.  An initial $\nue$ will be
removed from the beam and the resulting electron very quickly brought
to rest.  With an initial $\numu$, the produced muon will eventually
decay.  However, the laboratory-frame decay length for the muon is
always much longer than its range, the distance over which its
electromagnetic energy losses bring it to rest.  Thus the neutrinos
from its decay will be below 50 MeV, and of no further interest here.
The decay length and range for muons are shown in
Fig.~\ref{fig:lengths}.

As first noted by Ritz and Seckel \cite{Ritz}, the situation is very
different for $\nutau$, since at all but the highest energies, the tau
decay length is less than its range, so that the tau will decay in
flight without significant energy loss; see Fig.\ \ref{fig:lengths}.  Tau
decays occur by many branches, but all contain a $\nutau$.  This
regenerated $\nutau$ carries a fraction about $0.75 \times 0.4 \simeq
1/3$ of the initial $\nutau$ energy, where 0.4 is for $\nutau$ from
tau decay \cite{DRS,Gaisserbook}.  Ritz and Seckel considered WIMP
annihilation in the sun, and the propagation of the produced
neutrinos.  The utility of this regeneration effect for the detection
of astrophysical neutrinos was not noted until a decade later, by
Halzen and Saltzberg~\cite{Halzen}.

%%%%%%%%%%%%%%%%%%%%%%%%%%%%%%%%%%%%%%%%%%%%%%%%%%%%%%%%%%%%%%%%%%%%%%
%%%%%%%%%%%%%%%%%%%%%%%%%%%%%%%%%%%%%%%%%%%%%%%%%%%%%%%%%%%%%%%%%%%%%%

The regeneration process will continue (along with the less frequent
NC scatterings) until the $\nutau$ energy has been moderated to the
energy such that the neutrino interaction length is comparable to the
remaining distance in Earth.  At nadir angle $\psi = 0^\circ$, such
that the neutrino will cross the full diameter of Earth, the
transparency energy is about 40 TeV (at larger nadir angles, neutrinos
at higher energies can pass through without scattering).  The
distribution of $\nutau$ energies around the transparency energy is
characteristically lognormal, with a one-sigma width of approximately
one decade.  Thus all of the initial $\nutau$ will emerge at
relatively high energies.

Future km-scale astrophysical neutrino detectors are designed for the
detection of $\nue$ (via an electromagnetic shower) or $\numu$ (via
the track of a penetrating muon).  The detection of $\nutau$ is much
more difficult.  Learned and Pakvasa~\cite{doublebang} have shown that
$\nutau$ may be detected by a ``double-bang'' signature, if the tau
lepton production and decay, each accompanied by a large shower, are
well-separated but both occur in the detector.  The double-bang events
will be observable only in a narrow range around several PeV, and only
at large nadir angles, so other techniques for $\nutau$ detection are
needed.

This is precisely the point of the signal proposed by Halzen and
Saltzberg~\cite{Halzen}.  In their scheme, the $\nutau$ interacts
outside the detector and the tau created decays as $\tau^- \rightarrow
\mu^- \nubmu \nutau$, where the muon, which has a long range, is seen
in the detector.  These $\nutau$ events can be separated from $\numu$
CC events (which also produce a muon) on a statistical basis, using
the fact that the $\nutau$ nadir angle distribution will be unchanged
by passage through Earth, whereas that for $\numu$ will show
exponential absorption (as will $\nue$).  Additionally, if the
incoming neutrino spectrum is shallower than about $E^{-2}$, the
pileup near the transparency energy may be visible~\cite{Halzen,DRS}.
However, it should be noted that the above tau branching ratio of 18\%
reduces the detectability of the $\nutau$ flux compared to a $\numu$
flux at the same energy, both taken just below the detector.  While
the muon in the $\nutau$-induced case will have an energy roughly 0.4
of that in the $\numu$-induced case because of the tau
decay~\cite{DRS,Gaisserbook}, the muon range and hence detectability
varies only slowly with energy (see Fig~\ref{fig:lengths}).

%%%%%%%%%%%%%%%%%%%%%%%%%%%%%%%%%%%%%%%%%%%%%%%%%%%%%%%%%%%%%%%%%%%%%%
%%%%%%%%%%%%%%%%%%%%%%%%%%%%%%%%%%%%%%%%%%%%%%%%%%%%%%%%%%%%%%%%%%%%%%
%\section{Secondary antineutrinos}
%%%%%%%%%%%%%%%%%%%%%%%%%%%%%%%%%%%%%%%%%%%%%%%%%%%%%%%%%%%%%%%%%%%%%%
%%%%%%%%%%%%%%%%%%%%%%%%%%%%%%%%%%%%%%%%%%%%%%%%%%%%%%%%%%%%%%%%%%%%%%

{\it Secondary antineutrinos.---}
While expected to be a significant component of the astrophysical
neutrino flux, and the only component which is not attenuated in
Earth, $\nutau$ are not easily detected.  However, we point out a new
effect that significantly improves their detectability.  In each step
of the regeneration process $\nutau \rightarrow \tau \rightarrow
\nutau$ described above, the tau lepton decay will always produce a
$\nutau$.  Additionally, 18\% of decays are $\tau \rightarrow \nutau
\mu \nubmu$ and 18\% are $\tau \rightarrow \nutau e \nube$.  In the
laboratory frame, on average $\nutau$ and $\mu$ each carry a fraction
0.4 of the tau energy and $\nubmu$ carries a fraction
0.2~\cite{DRS,Gaisserbook}.  Secondary $\numu$ and $\nue$ are also
produced by initial $\nubtau$ (expected to have a flux comparable to
the initial $\nutau$).  These secondary antineutrinos ($\nube,
\nubmu$) and neutrinos ($\nue, \numu$) have not been taken into
account before, but they have a surprisingly large effect on the
detectability of $\nutau$ and $\nubtau$.

The number of regeneration steps for $\nutau$ is approximately $N =
\log(E_i/E_T)/\log{3}$, where $E_i$ is the initial energy and $E_T$ is
the transparency energy.  With more steps, there are more
opportunities to create secondary $\nubmu$ and $\nube$.  However,
those created farther from the detector have a greater chance of being
absorbed.  On each step the secondary antineutrinos will carry about
1/6 of the initial $\nutau$ energy, reflecting the energy lost in the
CC scattering and the tau decay.

It is generally assumed that all scatterings and decays are collinear.
However, as the number of regeneration steps increases, the cumulative
non-collinearity could be large enough to blur astrophysical point
sources.  If the initial neutrino has energy $E_\nu$, simple
considerations indicate that for $\nutau \rightarrow \tau$, the
scattering angle is about $1^\circ/\sqrt{E_\nu/{\rm\ TeV}}$, and for
$\tau \rightarrow \nutau$, the decay angle is about
$0.2^\circ/(E_\nu/{\rm\ TeV})$.  The decay angle can thus be neglected
compared to the scattering angle.  In order for the number of
scatterings to be nonzero, $E_\nu$ must be above about 100 TeV, so
that on the last step the angular deviation is about $0.1^\circ$.
Taking previous regeneration steps into account gives a maximum
deviation of about $0.3^\circ$.  This is below the expected $1^\circ$
reconstruction resolution of the proposed detectors~\cite{Learned},
and so astrophysical point sources should not be blurred by $\nutau$
regeneration.

%%%%%%%%%%%%%%%%%%%%%%%%%%%%%%%%%%%%%%%%%%%%%%%%%%%%%%%%%%%%%%%%%%%%%%
%%%%%%%%%%%%%%%%%%%%%%%%%%%%%%%%%%%%%%%%%%%%%%%%%%%%%%%%%%%%%%%%%%%%%%
%\subsection{Analytic estimate of secondary antineutrinos}
%%%%%%%%%%%%%%%%%%%%%%%%%%%%%%%%%%%%%%%%%%%%%%%%%%%%%%%%%%%%%%%%%%%%%%
%%%%%%%%%%%%%%%%%%%%%%%%%%%%%%%%%%%%%%%%%%%%%%%%%%%%%%%%%%%%%%%%%%%%%%

We can estimate the secondary $\nube$ and $\nubmu$ number flux
produced by a $\nutau$ beam by making the following crude assumptions:
(i) $\nutau$ do not lose energy in their CC or NC interactions; (ii)
$\nube$ or $\nubmu$ which interact are removed from the beam; and
(iii) $\nube$ or $\nubmu$ from tau decay are produced with the same
energy as the initial $\nutau$.  Consider a beam of astrophysical
neutrinos in the expected ratio $\nue:\numu:\nutau = 1:1:1$, each with
a flux $F_0$, and all at an energy $E_0$.  Define $f$ as the fraction
of $\nutau$ interactions that produce a a $\nube$ or $\nubmu$: $f =
0.71 \times 0.18 = 0.13$, where the first factor is the CC/NC
fraction, and the second factor is the relevant tau-decay branching
ratio.  The secondary $\nubmu$ flux from tau decay obeys the equation
\begin{displaymath}
\frac{d}{d x} F_\nubmu(x) = 
\frac{f}{\lambda}F_0 - \frac{1}{\lambda} F_\nubmu(x) \,,
\end{displaymath}
where $x$ is the distance through Earth at the chosen nadir angle.
The first term represents the creation of $\nubmu$ from the $\nutau$
flux, and the second represents their subsequent absorption.  Note
that the interaction length $\lambda$ is the same for both flavors.
Thus, in this simple estimate, the fluxes after propagation through
Earth are
\begin{eqnarray}
F_\nue/F_0 & = & F_\numu/F_0 = e^{-x/\lambda}\,; \
F_\nutau/F_0 = 1 \nonumber \\
F_\nube/F_0 & = & F_\nubmu/F_0 = f \left(1 - e^{-x/\lambda}\right)\,; \
F_\nubtau/F_0 = 0\,. \nonumber
\end{eqnarray}
The initial $\nue$ and $\numu$ fluxes are exponentially depleted.
However, the flux associated with the initial $\nutau$ is increased
due to the secondary $\nube$ and $\nubmu$.  For large $x$,
corresponding to more $\nutau$ regeneration steps, the increased
production of $\nube$ and $\nubmu$ is balanced by their increased
absorption.  For the expected initial antineutrinos (with the same
fluxes and flavor ratios as for neutrinos), similar results obtain for
the production of secondary neutrinos, and they must be added to the
above.  For small $x$, the initial $\nue$, $\numu$, $\nube$, $\nubmu$
fluxes will survive, whereas for large $x$, they will be absorbed but
replaced by the approximately constant secondary fluxes created by
$\nutau$ and $\nubtau$.  Thus even for large energies and/or small
nadir angles, the fluxes of $\nue$, $\numu$, $\nube$, $\nubmu$ {\it
never vanish}, contrary to previous predictions.

%%%%%%%%%%%%%%%%%%%%%%%%%%%%%%%%%%%%%%%%%%%%%%%%%%%%%%%%%%%%%%%%%%%%%%
%%%%%%%%%%%%%%%%%%%%%%%%%%%%%%%%%%%%%%%%%%%%%%%%%%%%%%%%%%%%%%%%%%%%%%
%\subsection{Numerical calculation of secondary antineutrinos}
%%%%%%%%%%%%%%%%%%%%%%%%%%%%%%%%%%%%%%%%%%%%%%%%%%%%%%%%%%%%%%%%%%%%%%
%%%%%%%%%%%%%%%%%%%%%%%%%%%%%%%%%%%%%%%%%%%%%%%%%%%%%%%%%%%%%%%%%%%%%%

We have calculated the secondary $\nube$ and $\nubmu$ production with
a Monte Carlo code that simulates the passage of high-energy neutrinos
through Earth.  We assume collinear propagation for all particles
followed, as justified above.  The code starts with an initial
neutrino and randomly samples its interaction probability at each step
in distance, using the CC and NC total cross sections from Gandhi,
Quigg, Reno, and Sarcevic~\cite{GQRS}.  In each interaction, we sample
the outgoing charged lepton or neutrino energy around the average
inelasticity values $\langle y \rangle$ given in Ref.~\cite{GQRS}.  At
these energies, $\langle y \rangle$ is the same for CC or NC
interactions, and the same for neutrinos or antineutrinos.
Accordingly, for the distributions about $\langle y \rangle$ we use
the $d\sigma/dy$ distribution given in Ref.~\cite{GQRS} for
neutrino-nucleon CC scattering.

After a NC interaction, the neutrino is followed until the next
interaction.  After a CC interaction, if a muon or tau is created, it
is followed.  For their energy loss, we use the results of
Ref.~\cite{DRSS}, and for their decays we use the laboratory-frame
decay probability distributions.  Any electrons created are assumed
stopped.  For the laboratory-frame $\nutau$ distribution from tau
decays, summed over all branches and including the tau polarization,
we use the distributions given in Ref.~\cite{DRS,Gaisserbook}.  In tau
decays leading to $\nube$ or $\nubmu$, we use the distributions for
those branches~\cite{DRS,Gaisserbook}.  There are many tau decay
branches that yield charged pions or kaons.  However, these mesons
will be stopped before they decay, so that the produced neutrinos are
at low energies and hence ignored.  For all particles, we impose a
low-energy cutoff at 100 GeV, though the results are insensitive to
the cutoff as long as it is well below the transparency energy.

\begin{figure}[t]
\centerline{\epsfxsize=3.5in \epsfbox{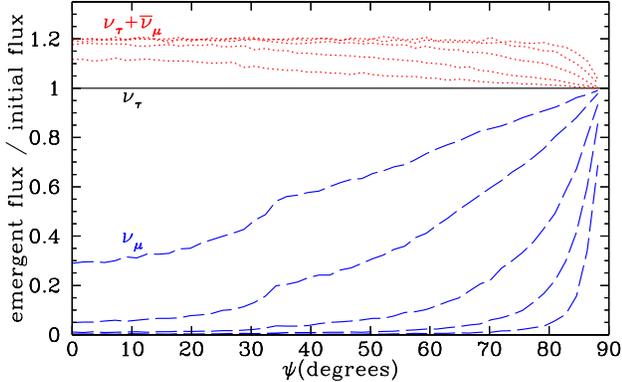}}
\caption{The surviving fluxes associated with the initial $\nutau$ and
$\numu$ fluxes versus nadir angle, for a variety of energies.  The
results were calculated with our Monte Carlo code, and the slight
jaggedness in the curves is due to statistical fluctuations.  The
initial $\numu$ flux is depleted ($\nue$ would be the
same), and the flux associated with the initial $\nutau$ is increased
due to the secondary $\nubmu$ (and the equal $\nube$ component,
not shown).  The initial energies are $10^5$, $10^6$, $10^7$, $10^8$,
and $10^9$ GeV, going from top to bottom for the $\numu$ fluxes and
bottom to top for the $\nutau + \nubmu$ fluxes.  The emergent $\nutau$
flux is unity for all energies.  We use the density profile given in
the Preliminary Reference Earth Model described in, e.g.,
Ref.~\protect\cite{GQRS}.  Note the mantle-core transition at $\psi
\simeq 33^\circ$, corresponding to a radial distance of 3480 km.}
\label{fig:nadir-num}
\end{figure}

In Fig.~\ref{fig:nadir-num} (cf. Fig.~2 of Ref.~\cite{Halzen}), we
plot the numerical results for the emergent $\numu$ and $\nutau +
\nubmu$ fluxes using several initial energies.  A very small fraction
(about 1\%) of $\nutau$ either downscatter to very low energies by
repeated NC interactions, or the tau lepton emerges from Earth; these
are neglected in Fig.~\ref{fig:nadir-num}.  The results are in general
agreement with the analytic estimates, though the $\nubmu$ flux is
about twice as large as the analytic estimate.  The principal reason
for this is the fact that the secondary $\nubmu$ are about 1/2 as
energetic as the regenerated $\nutau$, and so have a longer
interaction length than assumed in the analytic estimate.  They are
created in the last few interaction lengths of the $\nutau$.  We find
that the $\nutau$ and $\nubmu$ are distributed around the transparency
energy, as expected.  The surviving $\numu$ have the initial energy
$E_0$, at least until their flux is greatly suppressed.  A very small
fraction will be moderated down to the transparency energy by repeated
NC interactions; this is why the $\numu$ flux in
Fig.~\ref{fig:nadir-num} never completely vanishes.

%%%%%%%%%%%%%%%%%%%%%%%%%%%%%%%%%%%%%%%%%%%%%%%%%%%%%%%%%%%%%%%%%%%%%%
%%%%%%%%%%%%%%%%%%%%%%%%%%%%%%%%%%%%%%%%%%%%%%%%%%%%%%%%%%%%%%%%%%%%%%

%%%%%%%%%%%%%%%%%%%%%%%%%%%%%%%%%%%%%%%%%%%%%%%%%%%%%%%%%%%%%%%%%%%%%%
%%%%%%%%%%%%%%%%%%%%%%%%%%%%%%%%%%%%%%%%%%%%%%%%%%%%%%%%%%%%%%%%%%%%%%
%\section{Conclusions}
%%%%%%%%%%%%%%%%%%%%%%%%%%%%%%%%%%%%%%%%%%%%%%%%%%%%%%%%%%%%%%%%%%%%%%
%%%%%%%%%%%%%%%%%%%%%%%%%%%%%%%%%%%%%%%%%%%%%%%%%%%%%%%%%%%%%%%%%%%%%%

{\it Conclusions.---}
Earth absorbs high-energy $\nue$ and $\numu$.  In fact, high-energy
$\nutau$ are absorbed too, but the $\nutau$ flux is regenerated by
prompt tau decays, thus moderating the $\nutau$ energies down to near
the transparency energy.  As a function of the nadir angle, the $\nue$
and $\numu$ fluxes are exponentially absorbed, while the $\nutau$ flux
remains unchanged by passage through Earth.  Halzen and Saltzberg have
proposed that these $\nutau$ can be detected by CC interactions after
which the tau decays to a muon below the detector, and that they can
be separated from $\numu$ CC interactions by their characteristic
nadir-angle dependence~\cite{Halzen}.  A difficulty with this
technique is the 18\% branching fraction for the tau decay into a
muon.

We have pointed out a new effect, that the $\nutau \rightarrow \tau
\rightarrow \nutau$ regeneration process creates a secondary $\nubmu$
flux.  Though their flux is at most 0.2 of the $\nutau$ flux, they are
as detectable by the production of muons as the entire $\nutau$ flux.
Similarly, a secondary $\nube$ flux is created that doubles the
detectability of $\nutau$ by the production of electrons and their
associated showers.  The detectability of $\nutau$ by neutral-current
channels will be about 40\% larger.  For astrophysical antineutrinos,
the secondary neutrinos created have the same effect, again doubling
the detectability of the $\nubtau$ flux.  Taking these secondary
fluxes into account in the energy and nadir-angle distributions (e.g.,
in studies like Refs.~\cite{DRS,others}) will be essential for
understanding the spectra and flavor composition of astrophysical
neutrinos.

%%%%%%%%%%%%%%%%%%%%%%%%%%%%%%%%%%%%%%%%%%%%%%%%%%%%%%%%%%%%%%%%%%%%%%
%%%%%%%%%%%%%%%%%%%%%%%%%%%%%%%%%%%%%%%%%%%%%%%%%%%%%%%%%%%%%%%%%%%%%%
%\section*{Acknowledgements}
%%%%%%%%%%%%%%%%%%%%%%%%%%%%%%%%%%%%%%%%%%%%%%%%%%%%%%%%%%%%%%%%%%%%%%
%%%%%%%%%%%%%%%%%%%%%%%%%%%%%%%%%%%%%%%%%%%%%%%%%%%%%%%%%%%%%%%%%%%%%%

We thank Francis Halzen, Dan Hooper, David Saltzberg, and Ina Sarcevic
for useful discussions.  J.F.B. (as a David N. Schramm Fellow) and
E.W.K. were supported by NASA under NAG5-10842, and by Fermilab, which
is operated by URA under DOE contract No.\ DE-AC02-76CH03000.
E.W.K. thanks the Korea Institute for Advanced Study for their
hospitality while this work was completed.  P.C. was supported by DOE
grant No.  5-90098.

\vspace{-0.5cm}

%%%%%%%%%%%%%%%%%%%%%%%%%%%%%%%%%%%%%%%%%%%%%%%%%%%%%%%%%%%%%%%%%%%%%%
%%%%%%%%%%%%%%%%%%%%%%%%%%%%%%%%%%%%%%%%%%%%%%%%%%%%%%%%%%%%%%%%%%%%%%

%%%%%%%%%%%%%%%%%%%%%%%%%%%%%%%%%%%%%%%%%%%%%%%%%%%%%%%%%%%%%%%%%%%%%%
%%%%%%%%%%%%%%%%%%%%%%%%%%%%%%%%%%%%%%%%%%%%%%%%%%%%%%%%%%%%%%%%%%%%%%


\begin{references}
\frenchspacing
%%%%%%%%%%%%%%%%%%%%%%%%%%%%%%%%%%%%%%%%%%%%%%%%%%%%%%%%%%%%%%%%%%%%%%
%%%%%%%%%%%%%%%%%%%%%%%%%%%%%%%%%%%%%%%%%%%%%%%%%%%%%%%%%%%%%%%%%%%%%%

\vspace{-1.5cm}

\bibitem{Gaisser}
T.~K.~Gaisser, F.~Halzen and T.~Stanev, 
Phys.\ Rept.\  {\bf 258}, 173 (1995) [Erratum-ibid.\ {\bf 271}, 355 (1995)].
%%CITATION = HEP-PH 9410384;%%

\bibitem{Learned}
J.~G.~Learned and K.~Mannheim,
Ann.\ Rev.\ Nucl.\ Part.\ Sci.\ {\bf 50}, 679 (2000).
%%CITATION = ARNUA,50,679;%%

\bibitem{Pasquali}
L.~Pasquali, M.~H.~Reno and I.~Sarcevic, hep-ph/9905389.
%%CITATION = HEP-PH 9905389;%%

\bibitem{SK}
Y.~Fukuda {\it et al.}, Phys.\ Rev.\ Lett.\  {\bf 81}, 1562 (1998);
%%CITATION = HEP-EX 9807003;%%
S.~Fukuda {\it et al.}, Phys.\ Rev.\ Lett.\  {\bf 85}, 3999 (2000).
%%CITATION = HEP-EX 0009001;%%

\bibitem{AHK}
I.~F.~Albuquerque, L.~Hui and E.~W.~Kolb,
Phys.\ Rev.\ D {\bf 64}, 083504 (2001).
%%CITATION = HEP-PH 0009017;%%

\bibitem{MacGibbon}
J.~H.~MacGibbon, U.~F.~Wichoski and B.~R.~Webber, hep-ph/0106337.
%%CITATION = HEP-PH 0106337;%%

\bibitem{Halzen}
F.~Halzen and D.~Saltzberg, Phys.\ Rev.\ Lett.\ {\bf 81}, 4305 (1998).
%%CITATION = HEP-PH 9804354;%%

\bibitem{GQRS}
R.~Gandhi, C.~Quigg, M.~H.~Reno and I.~Sarcevic,
Astropart.\ Phys.\  {\bf 5}, 81 (1996);
%%CITATION = HEP-PH 9512364;%%
Phys.\ Rev.\ D {\bf 58}, 093009 (1998).
%%CITATION = HEP-PH 9807264;%%

\bibitem{Ritz}
S.~Ritz and D.~Seckel, Nucl.\ Phys.\ B {\bf 304}, 877 (1988).
%%CITATION = NUPHA,B304,877;%%

\bibitem{DRS}
S.~I.~Dutta, M.~H.~Reno and I.~Sarcevic,
Phys.\ Rev.\ D {\bf 62}, 123001 (2000).
%%CITATION = HEP-PH 0005310;%%

\bibitem{Gaisserbook}
T.~K. Gaisser, {\it Cosmic Rays and Particle Physics},
(Cambridge Univ. Press, Cambridge, 1992).

\bibitem{doublebang}
J.~G.~Learned and S.~Pakvasa, Astropart.\ Phys.\  {\bf 3}, 267 (1995).
%%CITATION = HEP-PH 9405296;%%

\bibitem{DRSS}
S.~I.~Dutta, M.~H.~Reno, I.~Sarcevic and D.~Seckel,
Phys.\ Rev.\ D {\bf 63}, 094020 (2001).
%%CITATION = HEP-PH 0012350;%%

\bibitem{others}
P.~Jain, J.~P.~Ralston and G.~M.~Frichter,
Astropart.\ Phys.\  {\bf 12}, 193 (1999);
%%CITATION = HEP-PH 9902206;%%
F.~Becattini and S.~Bottai, Astropart.\ Phys.\  {\bf 15}, 323 (2001);
%%CITATION = ASTRO-PH 0003179;%%
H.~Athar, M.~Jezabek and O.~Yasuda, 
Phys.\ Rev.\ D {\bf 62}, 103007 (2000);
%%CITATION = HEP-PH 0005104;%%
I.~F.~Albuquerque, J.~Lamoureux and G.~F.~Smoot, hep-ph/0109177.
%%CITATION = HEP-PH 0109177;%%

%%%%%%%%%%%%%%%%%%%%%%%%%%%%%%%%%%%%%%%%%%%%%%%%%%%%%%%%%%%%%%%%%%%%%%
%%%%%%%%%%%%%%%%%%%%%%%%%%%%%%%%%%%%%%%%%%%%%%%%%%%%%%%%%%%%%%%%%%%%%%
\end{references}
\end{document}